\documentclass[
reprint,
twocolumn,
superscriptaddress,
amsmath,amssymb,
aps,
prl,
]{revtex4-2}
\usepackage{amsmath}
\usepackage{amssymb}
\usepackage{amsfonts}
\usepackage{graphicx}
\usepackage{mathrsfs}
\usepackage{mathtools}
\usepackage{multirow}
\usepackage{siunitx}
\usepackage[modulo]{lineno}
\usepackage{soul}
\usepackage[normalem]{ulem}
\usepackage{booktabs}
\usepackage{float}
\usepackage[colorlinks=true,allcolors=blue]{hyperref}
\usepackage{enumitem}
\usepackage{color}
\usepackage{lineno}
\usepackage{xcolor}
\usepackage{comment}
\usepackage{subfigure}
\newlist{todolist}{itemize}{2}
\setlist[todolist]{label=$\square$}
\usepackage{pifont}
\usepackage{orcidlink}

\begin{document}

\title{Precise $^{136}$Xe Double Beta Decay Measurement in PandaX-4T with Implications on the Nuclear Matrix Elements and Majorons}

% !TEX root = ../main.
%updated on 2025/06/09

\def\tdli{State Key Laboratory of Dark Matter Physics, Key Laboratory for Particle Astrophysics and Cosmology (MoE), Shanghai Key Laboratory for Particle Physics and Cosmology, Tsung-Dao Lee Institute \& School of Physics and Astronomy, Shanghai Jiao Tong University, Shanghai 201210, China}
\def\sjtuphys{State Key Laboratory of Dark Matter Physics, Key Laboratory for Particle Astrophysics and Cosmology (MoE), Shanghai Key Laboratory for Particle Physics and Cosmology, School of Physics and Astronomy, Shanghai Jiao Tong University, Shanghai 200240, China}
\def\newcorner{New Cornerstone Science Laboratory, Tsung-Dao Lee Institute, Shanghai Jiao Tong University, Shanghai 201210, China}
\def\MESJTU{School of Mechanical Engineering, Shanghai Jiao Tong University, Shanghai 200240, China}
\def\SPEIT{SJTU Paris Elite Institute of Technology, Shanghai Jiao Tong University, Shanghai 200240, China}
\def\SJTUSC{Shanghai Jiao Tong University Sichuan Research Institute, Chengdu 610213, China}

\def\BUAA{School of Physics, Beihang University, Beijing 102206, China}
\def\BUAACenter{Peng Huanwu Collaborative Center for Research and Education, Beihang University, Beijing 100191, China}
\def\BUAALab{International Research Center for Nuclei and Particles in the Cosmos \& Beijing Key Laboratory of Advanced Nuclear Materials and Physics, Beihang University, Beijing 100191, China}
\def\SCNT{Southern Center for Nuclear-Science Theory (SCNT), Institute of Modern Physics, Chinese Academy of Sciences, Huizhou 516000, China}

\def\USTClab{State Key Laboratory of Particle Detection and Electronics, University of Science and Technology of China, Hefei 230026, China}
\def\USTCdep{Department of Modern Physics, University of Science and Technology of China, Hefei 230026, China}

\def\YaLongSD{Yalong River Hydropower Development Company, Ltd., 288 Shuanglin Road, Chengdu 610051, China}
\def\scKeyLab{Jinping Deep Underground Frontier Science and Dark Matter Key Laboratory of Sichuan Province, Liangshan 615000, China}

\def\pku{School of Physics, Peking University, Beijing 100871, China}
\def\CHEPpku{Center for High Energy Physics, Peking University, Beijing 100871, China}

\def\SDUdep{Research Center for Particle Science and Technology, Institute of Frontier and Interdisciplinary Science, Shandong University, Qingdao 266237, China}
\def\SDUlab{Key Laboratory of Particle Physics and Particle Irradiation of Ministry of Education, Shandong University, Qingdao 266237, China}

\def\UMD{Department of Physics, University of Maryland, College Park, Maryland 20742, USA}

\def\SYU{School of Physics, Sun Yat-Sen University, Guangzhou 510275, China}
\def\SYUSFI{Sino-French Institute of Nuclear Engineering and Technology, Sun Yat-Sen University, Zhuhai 519082, China}
\def\SYUzhuhai{School of Physics and Astronomy, Sun Yat-Sen University, Zhuhai 519082, China}
\def\SYUshenzhen{School of Science, Sun Yat-Sen University, Shenzhen 518107, China}

\def\NKU{School of Physics, Nankai University, Tianjin 300071, China}
\def\YTU{Department of Physics, Yantai University, Yantai 264005, China}
\def\FDU{Key Laboratory of Nuclear Physics and Ion-beam Application (MOE), Institute of Modern Physics, Fudan University, Shanghai 200433, China}
\def\CDUT{College of Nuclear Technology and Automation Engineering, Chengdu University of Technology, Chengdu 610059, China}

\def\CASIMP{Institute of Modern Physics, Chinese Academy of Sciences, Lanzhou 730000, China}
\def\UCASSNST{School of Nuclear Science and Technology, University of Chinese Academy of Sciences, Beijing 100049, China}
\def\CASIHEP{Institute of High Energy Physics, Chinese Academy of Sciences, Beijing 100049, China}
\def\UCASSPS{School of Physical Sciences, University of Chinese Academy of Sciences, Beijing 100049, China}

\affiliation{\tdli}
\author{Zhe Yuan\orcidlink{0009-0008-5657-3584}}\affiliation{\FDU} 
\author{Zihao Bo\orcidlink{0009-0002-0743-5368}}\affiliation{\sjtuphys}
\author{Wei Chen\orcidlink{0009-0009-5911-7135}}\affiliation{\sjtuphys}
\author{Xun Chen\orcidlink{0000-0001-7961-7908}}\affiliation{\tdli}\affiliation{\SJTUSC}\affiliation{\scKeyLab}
\author{Yunhua Chen}\affiliation{\YaLongSD}\affiliation{\scKeyLab}
\author{Chen Cheng\orcidlink{0000-0003-0164-7538}}\affiliation{\BUAA}
\author{Xiangyi Cui}\affiliation{\tdli}
\author{Manna Deng}\affiliation{\SYUSFI}
\author{Yingjie Fan}\affiliation{\YTU}
\author{Deqing Fang}\affiliation{\FDU}
% ...
\collaboration{PandaX Collaboration}
\author{Dong-Liang Fang\orcidlink{0000-0002-9669-7994}}\affiliation{\CASIMP}\affiliation{\UCASSNST}
\author{Yu-Feng Li\orcidlink{0000-0002-2220-5248}}\affiliation{\CASIHEP}\affiliation{\UCASSPS}
%\noaffiliation

%\date{\today}
\begin{abstract}
The continuous spectrum of double beta decay ($\beta\beta$) provides a sensitive probe to test the predictions of the standard model and to search for signatures of new physics beyond it. 
We present a comprehensive analysis of the $^{136}$Xe $\beta\beta$ spectrum utilizing $39.1 \pm 0.7~\textrm{kg}\cdot\textrm{yr}$ of $^{136}$Xe exposure from the PandaX-4T experiment.
The analysis yields the most precise measurement to date of the $^{136}$Xe two-neutrino double beta decay ($2\nu\beta\beta$) half-life, $(2.14 \pm 0.05) \times 10^{21}$ years, the uncertainty of which is reduced by a factor of 2 compared to our previous result.
We measure the parameter $\xi_{31}^{2\nu}$, defined as the ratio between the subleading and leading components of the $^{136}$Xe $2\nu\beta\beta$ nuclear matrix element, to be $0.59^{+0.41}_{-0.38}$, which is consistent with theoretical predictions.
We also search for Majoron-emitting modes of $^{136}$Xe $\beta\beta$, establishing the most stringent limit for the spectral index $n=7$.
\end{abstract}

\maketitle

%\renewcommand{\linenumbersep}{3pt}
%\linenumbers

Double beta decay ($\beta\beta$) is a rare nuclear process~\cite{DBD:origin}. 
If two neutrinos are emitted during this process ($2\nu\beta\beta$), it is consistent with the electroweak interaction framework of the standard model of particle physics. 
In contrast, if no neutrinos are emitted, the process is referred to as neutrinoless double beta decay ($0\nu\beta\beta$), which can occur only if neutrinos are Majorana particles~\cite{NLDBD:probability,PhysRevD.25.2951,RevModPhys.95.025002,s40766-023-00049-2_revise}. 
Additionally, if one or two bosons are emitted in $\beta\beta$ processes, they are typically referred to as ``Majorons"~\cite{Majoron:goldstone,Majoron:Left-handed,Majoron:Unconventional}.
Although $2\nu\beta\beta$ has been experimentally observed and measured in several studies, such as those involving $^{136}$Xe~\cite{EXO-200:Xe136dbd2014,KamLAND-Zen:kesi31,NEXT:Xe136dbd,PandaX:2022kwg}, neither $0\nu\beta\beta$ nor Majoron-emitting $\beta\beta$ has been detected.

Recent experiments have established the most stringent lower limits on the $0\nu\beta\beta$ half-life of different isotopes, such as $^{136}$Xe~\cite{jkf6-48j8,EXO-200:2019nldbd}, $^{76}$Ge~\cite{GERDA:Genldbd2020,Majorana:2023Genldbd,CDEX:2024Genldbd,25tk-nctn}, and $^{130}$Te~\cite{science.adp6474}.
The $0\nu\beta\beta$ half-life is linked to the effective Majorana neutrino mass $m_{\beta\beta}$ by~\cite{s40766-023-00049-2_revise} 
\begin{equation}
 (T_{1/2}^{0\nu})^{-1} = \frac{|m_{\beta\beta}|^2}{m_{e}^2} g_{A}^4 |M^{0\nu}|^2 G^{0\nu},
\end{equation}
where $m_{e}$ is the electron mass, $g_{A}$ the axial-vector coupling constant, $M^{0\nu}$ the nuclear matrix element (NME), and $G^{0\nu}$ the phase space factor.
The current understanding of $0\nu\beta\beta$ NME remains incomplete, with significant discrepancies observed in different nuclear many-body approaches and potential influences of quenching effects~\cite{NME:Status,Xe136nldbd:2015review}.
Since $2\nu\beta\beta$ and $0\nu\beta\beta$ share the same initial and final nuclear states, accurately determining $2\nu\beta\beta$ NME is essential to constrain $0\nu\beta\beta$ NME predictions~\cite{kesi31:theory}.
A refined expression for the $2\nu\beta\beta$ half-life, retaining terms up to second order in the Taylor expansion, is presented as follows~\cite{kesi31:theory}:
\begin{equation}
\begin{aligned}
 (T_{1/2}^{2\nu})^{-1} = (g_{A}^{eff})^4 |M_\textrm{GT}^{2\nu}|^2
  (G_{0}^{2\nu}+\xi_{31}^{2\nu} G_{2}^{2\nu}),
\end{aligned}
\label{eq:NME}
\end{equation}
where $g_{A}^{eff}$ is the effective axial-vector coupling constant and $M_\textrm{GT}^{2\nu}$ denotes the leading-order Gamow-Teller component of the NME.
In contrast to the standard phase space factor $G_{0}^{2\nu}$, the phase space factor $G_{2}^{2\nu}$ accounts for the dependence on lepton energies of the energy denominators of NMEs.
The parameter $\xi_{31}^{2\nu}$ is defined as $\xi_{31}^{2\nu} = M_{\textrm{GT}-3}^{2\nu}/ M_\textrm{GT}^{2\nu}$, where the subleading NME $M_{\textrm{GT}-3}^{2\nu}$ is sensitive only to contributions from the lightest states of the intermediate nucleus, while the NME $M_\textrm{GT}^{2\nu}$ is also sensitive to higher-lying states.
The experimentally accessible parameter $\xi_{31}^{2\nu}$, which can be measured through electron energy spectrum fits, allows for the discrimination between different theoretical models, and thus, provides new insights into $0\nu\beta\beta$ NME calculations. 
Specifically, a zero (nonzero) value of $\xi_{31}^{2\nu}$ implies the higher (single) state dominance hypothesis of $2\nu\beta\beta$~\cite{kesi31:theory}.

Majorons are potential dark matter candidates and play roles in various cosmological and astrophysical processes~\cite{Majoron:darkmatter1,Majoron:darkmatter2,Majoron:darkmatter3}. 
The Majoron-emitting $\beta\beta$ encompasses a range of modes, which are distinguished by the number of emitted bosons, the leptonic charge, whether lepton number is violated, and whether the Majoron is a Goldstone boson. 
Different modes may produce identical summed electron energy spectra, which are classified into four types based on the spectral index $n =$ 1, 2, 3, and 7~\cite{Majoron:n137,Majoron:n2}.
Previous experimental searches have established stringent constraints on these processes with $^{136}$Xe~\cite{KamLAND-Zen:Majoron,EXO-200:Majoron2021}.

Previous measurements of $\xi_{31}^{2\nu}$ and searches for Majoron-emitting $\beta\beta$ of $^{136}$Xe have been limited to relatively high energy thresholds, specifically between 500~keV and 800~keV~\cite{KamLAND-Zen:kesi31,KamLAND-Zen:Majoron,EXO-200:Majoron2021}. 
Consequently, the lower energy region remains completely unexplored.
The PandaX-4T detector, which uses a natural xenon target and was initially designed for dark matter searches in the low-energy region, has expanded its detection capability to the MeV range~\cite{PandaX:2022kwg,PandaX-4T:Xe136nldbd}. 
In this Letter, we leverage the nearly complete $\beta\beta$ spectrum to perform the measurement of $\xi_{31}^{2\nu}$ and the searches for Majoron-emitting modes of $\beta\beta$ of $^{136}$Xe, using a dataset of the commissioning run (Run0, 94.8~days of live time) and the first science run (Run1, 163.5~days) of the PandaX-4T experiment~\cite{PandaX:2025WIMP,PandaX:SuperWIMP,PandaX-4T:Xe136nldbd} with a total $^{136}$Xe exposure of $39.1 \pm 0.7~\textrm{kg}\cdot\textrm{yr}$.
The energy region of interest (ROI) for the spectrum fit is from 20~keV to
2800~keV.

The PandaX-4T detector is a dual-phase time projection chamber (TPC) containing 3.7~tons of natural xenon in its active volume. 
The TPC has a vertical height of 1.185~m and is enclosed by a field cage along its lateral surfaces. 
The lateral surfaces of the TPC are lined with reflective Teflon panels, arranged in a regular 24-sided polygon with an inscribed diameter of 1.185~m.
Three electrode grids (cathode, gate, and anode) are vertically stacked within the TPC to generate a drift field and an extraction field. 
Two arrays of 3-inch Hamamatsu photomultiplier tubes (PMTs) are placed on the top and bottom as photosensors.
A more detailed description of the detector can be found in Ref.~\cite{PandaX-4T:2021bab}.

Energy deposition in liquid xenon (LXe) produces a prompt scintillation signal ($S1$) and ionization electrons. 
These electrons drift upward into the gaseous xenon phase, where they generate a delayed electroluminescence signal ($S2$) via proportional scintillation. 
Energy reconstruction utilizes the charge of $S1$ and $S2_\text{B}$, where $S2_\text{B}$ stands for $S2$ detected by the bottom PMT array, following the procedure detailed in Ref.~\cite{PandaX-4T:Xe136nldbd,PandaX:SuperWIMP,PandaX:2024med,PandaX:2023ggs}.
Spatial reconstruction determines the vertical ($z$-axis) position via drift velocity and the time delay between the $S1$ and $S2$ signals, while the horizontal ($x$- and $y$-axes) coordinates are obtained through maximum likelihood estimation of the charge distribution pattern of $S2$ observed in the top PMT array~\cite{PandaX:SuperWIMP}.

The data production and event selection procedures follow those described in Refs.~\cite{PandaX-4T:Xe136nldbd,PandaX:SuperWIMP,PandaX:2024med,PandaX:2023ggs}.
The fiducial volume (FV) from the previous study~\cite{PandaX:SuperWIMP}, defined as the innermost, cleanest part of the detector, is adopted in this analysis. 
The fiducial mass is determined to be $625 \pm 10$~kg for Run0 and $621 \pm 13$~kg for Run1, by scaling the ratio of $^{83\text{m}}\mathrm{Kr}$ events in the FV and in the whole TPC.
The uncertainty arises from the LXe density and the difference between the geometrically calculated volume and the volume scaled by the uniformly distributed $^{83\text{m}}\mathrm{Kr}$ calibration source~\cite{PandaX:SuperWIMP}.

Compared with our previous study of the $^{136}$Xe $2\nu\beta\beta$ decay~\cite{PandaX:2022kwg}, the present analysis extends the analysis window down to 20~keV. 
Similar to Ref.~\cite{PandaX:SuperWIMP}, monoenergetic peaks at 41.5~keV (from $^{83\text{m}}\mathrm{Kr}$), 164~keV ($^{131\text{m}}$Xe), and 236~keV ($^{127}$Xe and $^{129\text{m}}$Xe) are used for energy calibration, together with high-energy peaks from $^{137}$Cs, $^{60}$Co, $^{40}$K, and the $^{232}$Th decay chain.
The energy response is modeled using five parameters, which account for a possible shift between the reconstructed and true energies as well as the energy resolution~\cite{PandaX:SuperWIMP}. 
The energy resolution is parametrized by a Gaussian function with a width $\sigma(E)$ given by
$\frac{\sigma(E)}{E} = \frac{a}{\sqrt{E}} + bE + c$, where $E$ is the reconstructed energy in keV. The residual energy nonlinearity is modeled as $E = d\hat{E} + e$, where $\hat{E}$ denotes the true energy.
The calibrated parameter values and their uncertainties, $\mathcal{M}_{in} = {(a_{in}, b_{in}, c_{in}, d_{in}, e_{in})}^\mathrm{T}$ (Table~\ref{tab:syserr}), together with the corresponding $5 \times 5$ covariance matrix $\Sigma$, are used to fit the Run0 and Run1 data independently. 
Due to the lack of sufficient high-energy calibration data within the fiducial volume, the energy response model is derived from physics data in a control region outside the FV, providing a closely matched but statistically uncorrelated dataset for the final fit~\cite{PandaX-4T:Xe136nldbd}.

\begin{table}[tbp]
   \caption{
 Summary of sources of systematic uncertainties. 
   $\mathcal{M}_{in}$ denotes the 5-parameter detector response model (see text), with its means and uncertainties determined from monoenergetic peaks obtained from the control data outside of the FV.
 }
   \label{tab:syserr}
   \centering
   \renewcommand{\arraystretch}{1.5}
    \resizebox{\linewidth}{!}{
   \begin{tabular}{>{\centering\arraybackslash}m{.5cm}>{\centering\arraybackslash}m{.5cm}>{\centering\arraybackslash}m{.5cm}>{\centering\arraybackslash}m{.5cm}}
       \toprule
       \multicolumn{2}{c}{Sources} & \multicolumn{1}{c}{Run0 (94.8~d)} & \multicolumn{1}{c}{Run1 (163.5~d)} \\ 
       \hline
       \multicolumn{1}{c}{\multirow{5}{*}{\shortstack[t]{Detector \\ response}}}&\multicolumn{1}{c}{\multirow{1}{*}{\shortstack[t]{$a_{in}$~[$\sqrt{\mathrm{keV}}$]}}} & \multicolumn{1}{c}{$0.41 \pm 0.02$} & \multicolumn{1}{c}{$0.37 \pm 0.11$} \\
       \multicolumn{1}{c}{\multirow{2}{*}{}}&\multicolumn{1}{c}{\multirow{1}{*}{\shortstack[t]{$b_{in}$~[keV$^{-1}$]}}} & \multicolumn{1}{c}{$(5 \pm 1) \times 10^{-6}$} & \multicolumn{1}{c}{$(3 \pm 2) \times 10^{-6}$} \\
       \multicolumn{1}{c}{\multirow{2}{*}{}}&\multicolumn{1}{c}{\multirow{1}{*}{\shortstack[t]{$c_{in}$}}} & \multicolumn{1}{c}{$(4 \pm 20) \times 10^{-4}$} & \multicolumn{1}{c}{$(9 \pm 6) \times 10^{-3}$} \\
       \cline{2-4}
       \multicolumn{1}{c}{\multirow{2}{*}{}}&\multicolumn{1}{c}{\multirow{1}{*}{\shortstack[t]{$d_{in}$}}} & \multicolumn{1}{c}{$(1.0000 \pm 0.0006)$} & \multicolumn{1}{c}{$(1.0005 \pm 0.0006)$} \\
       \multicolumn{1}{c}{\multirow{2}{*}{}}&\multicolumn{1}{c}{\multirow{1}{*}{\shortstack[t]{$e_{in}$~[keV]}}} & \multicolumn{1}{c}{$(0.52 \pm 0.11)$} & \multicolumn{1}{c}{$(-3.12 \pm 0.75)$} \\
       \cline{1-4}
       \multicolumn{1}{c}{\multirow{2}{*}{\shortstack[t]{Overall \\ efficiency}}}&\multicolumn{1}{c}{SS fraction ($2\nu\beta\beta$)} & \multicolumn{1}{c}{($98 \pm 9$)\%} & \multicolumn{1}{c}{($98 \pm 10$)\%} \\
       \multicolumn{1}{c}{\multirow{2}{*}{}}&\multicolumn{1}{c}{Quality cut} & \multicolumn{1}{c}{($99.84\pm0.03$)\%} & \multicolumn{1}{c}{($99.72\pm0.11$)\%} \\
       \cline{1-4}
        \multicolumn{1}{c}{\multirow{3}{*}{\shortstack[t]{Signal \\ selection}}}&\multicolumn{1}{c}{LXe density~[g/cm$^{3}$]} & \multicolumn{2}{c}{$2.850\pm0.004$} \\
       \multicolumn{1}{c}{\multirow{3}{*}{}}&\multicolumn{1}{c}{$^{136}$Xe abundance} & \multicolumn{2}{c}{$(8.584 \pm 0.114)\%$} \\
       \multicolumn{1}{c}{\multirow{3}{*}{}}&\multicolumn{1}{c}{FV mass~[kg]} & \multicolumn{1}{c}{$625\pm10$} & \multicolumn{1}{c}{$621\pm13$} \\
       \toprule
   \end{tabular}
 }
\end{table}

The total detection efficiency comprises three components: the single-site (SS) fraction, quality cut (QC) efficiency, and ROI acceptance, estimated with methods inherited from Ref.~\cite{PandaX:SuperWIMP} and summarized in Table~\ref{tab:syserr}.
The SS fraction is defined as the ratio of SS to the sum of SS and multisite (MS) energy deposition events in the detector. 
We validate the SS fraction within the ROI using $^{232}$Th calibration data together with $^{232}$Th events simulated with the BambooMC simulation framework~\cite{Chen:2021asx}.
In this analysis, the QC efficiency of Run0 and Run1 are ($99.84 \pm 0.03$)\% and ($99.72 \pm 0.11$)\%, respectively.

To generate the signal spectra for the $\xi_{31}^{2\nu}$ fit, the theoretical spectra of the first and second terms of the Taylor expansion in Eq.~(\ref{eq:NME}) are calculated from the numerical wave functions, which are the solutions of the Dirac equation~\cite{NLDBDtheory:leftright}, obtained with the package RADIAL~\cite{Diracwave} by assuming a uniform distribution of the nuclear charge inside the nucleus.
The detector-response-convolved spectra are then simulated by the BambooMC package.
For $2\nu\beta\beta$ and Majoron-emitting $\beta\beta$ spectra, we use the Decay0 package~\cite{BxDecay0} to generate events with momenta of two electrons, and then, simulate the detector response with the BambooMC package as well.
\begin{figure}[h]
    \includegraphics[width=1.\columnwidth]{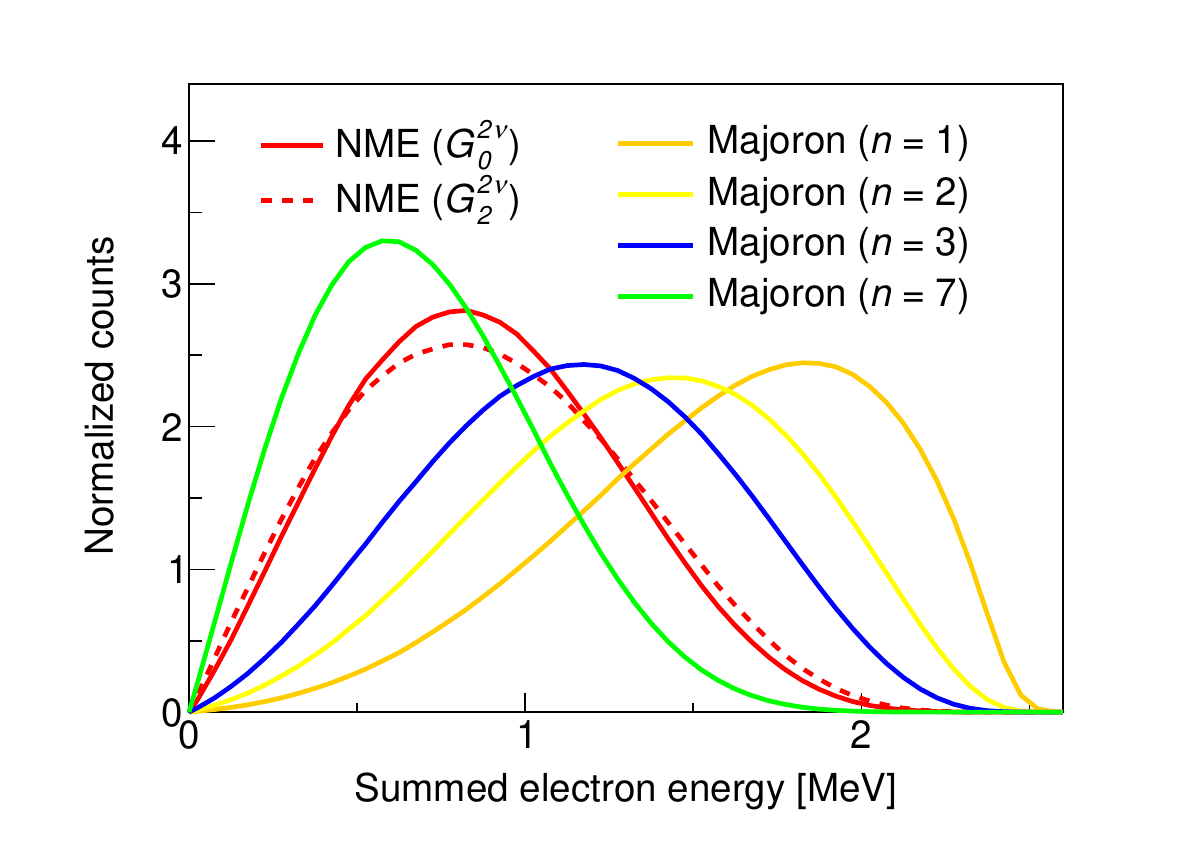}
    \caption{Spectra for $G_{0}^{2\nu}$, $G_{2}^{2\nu}$, and the Majoron-emitting $\beta\beta$ modes with $n$ = 1, 2, 3, and 7, respectively. Detector response model is taken into account.}
    \label{fig:Signal_spectra}
\end{figure}
The obtained spectra with detector response incorporated are shown in Fig.~\ref{fig:Signal_spectra}.
The ROI acceptance for $^{136}$Xe $2\nu\beta\beta$ in the $\xi_{31}^{2\nu}$ fit varies with $\xi_{31}^{2\nu}$, while the acceptances for Majoron-emitting modes are 100.00\% ($n=1$), 100.00\% ($n=2$), 99.99\% ($n=3$), and 99.95\% ($n=7$).

Background events originate from three primary sources: liquid xenon, the detector material and the stainless steel platform (SSP) material, and solar neutrinos, as summarized in Table~\ref{tab:bkg_summary}. 
$^{124}$Xe and $^{125}$I are taken from the measurements reported in Ref.~\cite{PandaX:2024xpq_revise}. 
For $^{127}$Xe, $^{129\mathrm{m}}$Xe, and $^{131\mathrm{m}}$Xe in Run0, we adopt the xenon-evolution analysis from Ref.~\cite{PandaX:SuperWIMP}.
In Run1, due to the absence of neutron calibration, the activities of these xenon isotopes are very low, and their time evolution cannot be determined precisely.
Therefore, we allow their activities to float in the fit.
The activities of $^{125}$Xe and $^{133}$Xe are left free in Run0 and are negligible in Run1.
$^{85}$Kr concentration is determined via $\beta$-$\gamma$ cascade tagging through the isomeric $^{85\mathrm{m}}$Rb~\cite{Collon:2004xs}. 
$^{214}$Pb and $^{212}$Pb components are also treated as free parameters. 
The material of the detector and the SSP contributes to the background.
Activities of $^{232}$Th, $^{238}$U, $^{60}$Co, and $^{40}$K in the detector components are taken from measurements using high-purity germanium counting stations~\cite{PandaX:materialHGe}, while activities of $^{232}$Th and $^{238}$U in SSP components are taken from Ref.~\cite{PandaX:SSP}. 
The solar $pp$ and $^{7}$Be neutrino contributions are taken from Refs.~\cite{Chen:2016eab,BOREXINO:2014pcl}.

\begin{table*}[tbp]
    \caption{The background contributions in the ROI for Run0 and Run1. 
 The fitted counts are obtained from the $^{136}$Xe $2\nu\beta\beta$ half-life measurement, the $\xi_{31}^{2\nu}$ measurement, and the Majoron-emitting $\beta\beta$ search with mode $n=7$.
 }
    \label{tab:bkg_summary}
    \centering
    \renewcommand{\arraystretch}{1.3}
    \begin{tabular}{>{\centering\arraybackslash}m{4cm}>{\centering\arraybackslash}m{3cm}>{\centering\arraybackslash}m{3cm}>{\centering\arraybackslash}m{3cm}>{\centering\arraybackslash}m{3cm}}
    \toprule
    \multirow{2}{*}{Components} & \multirow{2}{*}{Expected} & \multirow{2}{*}{$2\nu\beta\beta$} & \multirow{2}{*}{$\xi_{31}^{2\nu}$} & Majoron \\
     & & & & ($n=7$) \\
    \midrule
    $^{136}$Xe & \textemdash & $54848 \pm 1369$ & $55424 \pm 1400$ & $53558 \pm 1498$ \\
    \midrule
 SSP $^{232}$Th & $3813 \pm 343$ & $3433 \pm 235$ & $3439 \pm 227$ & $3425 \pm 237$ \\
 SSP $^{238}$U & $1672 \pm 502$ & $1884 \pm 373$ & $1683 \pm 384$ & $1931 \pm 368$ \\
 Detector $^{60}$Co & $3620 \pm 1810$ & $2671 \pm 163$ & $2627 \pm 160$ & $2708 \pm 173$ \\
 Detector $^{40}$K & $3840 \pm 1613$ & $2918 \pm 116$ & $2848 \pm 122$ & $2952 \pm 130$ \\
 Detector $^{232}$Th & $2942 \pm 1795$ & $4190 \pm 212$ & $4182 \pm 205$ & $4193 \pm 233$ \\
 Detector $^{238}$U & $2288 \pm 1396$ & $1844 \pm 252$ & $1786 \pm 230$ & $1869 \pm 245$ \\
    \midrule
 164~keV (Run0) & $40984 \pm 1978$ & $41635 \pm 403$ & $41633 \pm 389$ & $41627 \pm 1043$ \\
 208~keV (Run0) & $3643 \pm 154$ & $3858 \pm 76$ & $3854 \pm 75$ & $3846 \pm 110$ \\
 236~keV (Run0) & $54799 \pm 6660$ & $57306 \pm 530$ & $57305 \pm 511$ & $57301 \pm 1436$ \\
 380~keV (Run0) & $2450 \pm 177$ & $2408 \pm 71$ & $2405 \pm 69$ & $2409 \pm 86$ \\
 408~keV (Run0)  & $8567 \pm 406$ & $9167 \pm 125$ & $9162 \pm 124$ & $9165 \pm 238$ \\
    $^{125}$I(Run0) & $56 \pm 11$ & $61 \pm 10$ & $60 \pm 10$ & $60 \pm 10$ \\
    $^{125}$Xe (Run0) & float & $565 \pm 85$ & $569 \pm 81$ & $552 \pm 82$ \\
    $^{214}$Pb (Run0) & float & $11902 \pm 282$ & $11918 \pm 275$ & $11664 \pm 403$  \\
    $^{133}$Xe (Run0) & float & $8566 \pm 180$ & $8549 \pm 176$ & $8549 \pm 272$ \\
    $^{212}$Pb (Run0) & float & $1449 \pm 216$ & $1407 \pm 201$ & $1460 \pm 209$ \\
    $^{85}$Kr (Run0) & $469 \pm 244$ & $678 \pm 159$ & $659 \pm 155$ & $688 \pm 157$ \\
 164~keV (Run1) & float & $476 \pm 34$ & $473 \pm 35$ & $472 \pm 37$ \\  
 236~keV (Run1) & float & $301 \pm 33$ & $298 \pm 35$ & $296 \pm 36$ \\
    $^{125}$I (Run1) & $10 \pm 11$ & $9 \pm 9$ & $9 \pm 9$ & $9 \pm 9$ \\
    $^{214}$Pb (Run1) & float & $24586 \pm 561$ & $24588 \pm 558$ & $24389 \pm 739$ \\
    $^{212}$Pb (Run1) & float & $746 \pm 152$ & $705 \pm 144$ & $721 \pm 151$ \\
    $^{85}$Kr (Run1) & $1461 \pm 436$ & $2213 \pm 169$ & $2130 \pm 172$ & $2179 \pm 173$ \\
    $^{124}$Xe & $140 \pm 21$ & $137 \pm 13$ & $137 \pm 13$ & $136 \pm 13$ \\
    \midrule
    $pp+^{7}$Be $\nu$ & $196 \pm 21$ & $205 \pm 21$ & $205 \pm 20$ & $204 \pm 21$ \\
     \bottomrule
    \end{tabular}
\end{table*}

\begin{figure}[btp]
    \includegraphics[width=1.\columnwidth]{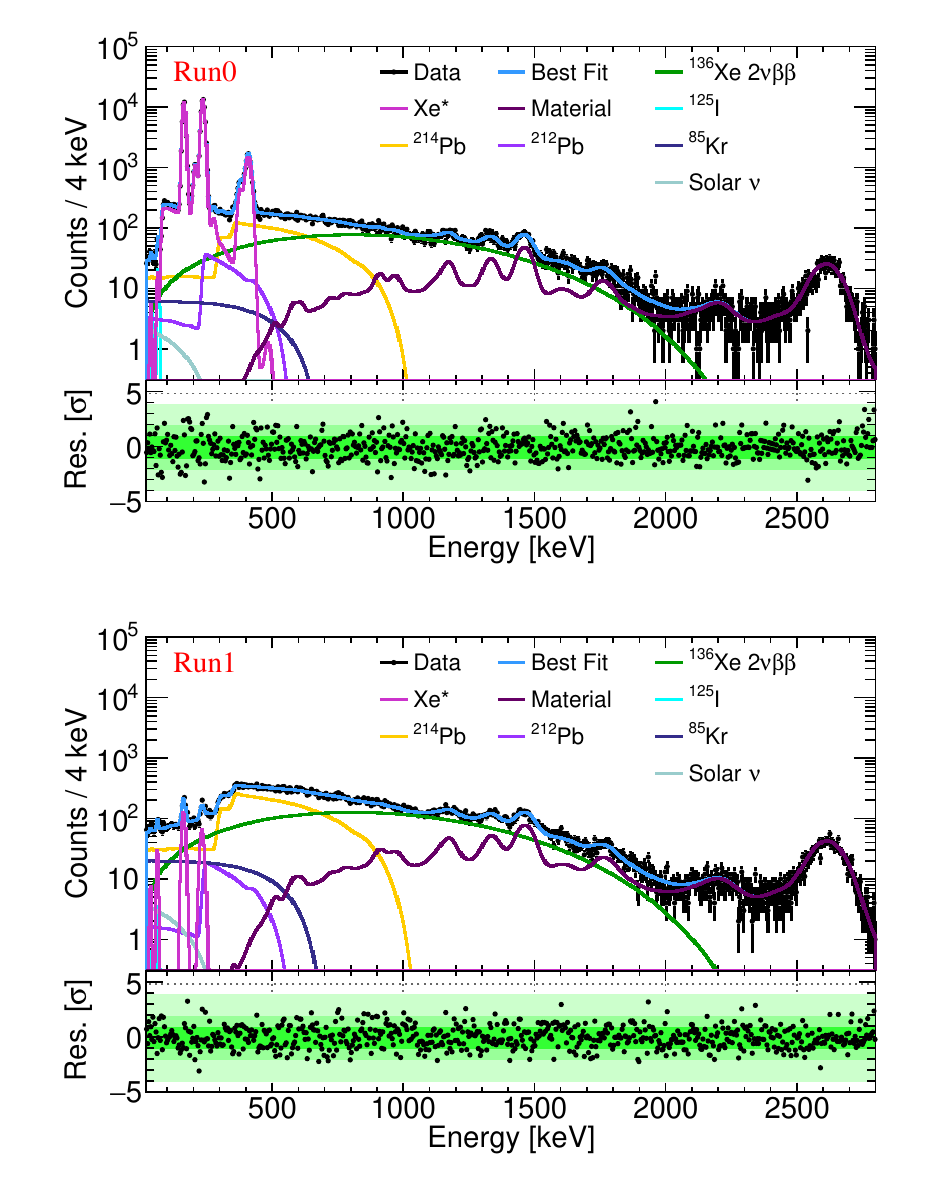}
    \caption{The SS data spectra and the fit for the $^{136}$Xe $2\nu\beta\beta$ half-life are shown for Run0 (top) and Run1 (bottom) from 20 keV to 2800 keV with a bin size of 4 keV. The horizontal axis represents the reconstructed energy in the data. Xe$^{*}$ denotes the contributions from $^{124}$Xe, $^{125}$Xe, $^{127}$Xe, $^{129m}$Xe, $^{131m}$Xe, and $^{133}$Xe. Material includes the contributions from SSP $^{232}$Th, SSP $^{238}$U, detector $^{60}$Co, detector $^{40}$K, detector $^{232}$Th, and detector $^{238}$U. The lower panel shows the residuals together with $\pm 1 \sigma$, $\pm 2 \sigma$, and $\pm 4 \sigma$ bands.
 }
    \label{fig:fitresult_DBD}
\end{figure}

Assuming no new physics is observed in this study, a one-dimensional binned likelihood fit is first performed to measure the half-life of $^{136}$Xe $2\nu\beta\beta$, using a Frequentist approach, as shown in Fig.~\ref{fig:fitresult_DBD}.
The likelihood is constructed as 
\begin{equation}
\begin{aligned}
 L = & \prod_{r=0}^{1} \prod_{i=1}^{N_\mathrm{bins}} \frac{(N_{r,i})^{N_{r,i}^\mathrm{obs}}e^{-N_{r,i}}}{N_{r,i}^\mathrm{obs}!}
 \mathcal{G}(\mathcal{M}_r; \mathcal{M}_r^{in}, \Sigma_r) \\
 & \times \prod_{j=1}^{N_\mathrm{G}} G(\eta_j; 0, \sigma_j),
\end{aligned}
\label{eq:likelihood}
\end{equation}
where $N_{r,i}$ and $N_{r,i}^\mathrm{obs}$ are the expected and observed numbers of events in the $i{\mathrm{th}}$ energy bin for Run-$r$, respectively.
$N_i$ in Run-$r$ is defined as
\begin{equation}
\begin{aligned}
 N_{i} = & (1+\eta_{a}) [ (1+\eta_s) n_{s} S_{i} \\
 & + \sum_{b=1}^{N_\mathrm{bkg}} (1+\eta_b) n_{b} B_{b, i}],
\end{aligned}
\label{eq:exp_count}
\end{equation}
where $n_s$ and $n_b$ are the counts of signal $s$ and background component $b$, respectively.
The corresponding $S_i$ and $B_{b,i}$ are the $i{\mathrm{th}}$ bin values of the normalized energy spectrum convolved with the five-parameter energy response model.
The Gaussian penalty term $\mathcal{G}(\mathcal{M}_r; \mathcal{M}_r^{in}, \Sigma_r)$ of the energy response contains the five-parameter $\mathcal{M}_r^{in}$ and the covariance matrix $\Sigma_r$ in Run-$r$.
The Gaussian penalty terms $G(\eta_j; 0, \sigma_j)$ are used to constrain the nuisance parameters $\eta_{a}$, $\eta_{s}$, and $\eta_{b}$, which represent the relative deviations in the overall efficiency, the signal selection, and the background model (see Table~\ref{tab:bkg_summary}), respectively.
The radioactivities of detector and SSP materials, $^{124}$Xe concentrations, and solar $pp$ and $^{7}$Be neutrino fluxes are identical in Run0 and Run1. 
Other background components are treated independently for Run0 and Run1.

The observed backgrounds are in general agreement with expectations, as shown in Table~\ref{tab:bkg_summary}. 
The fitted contributions from $^{\textrm{85}}$Kr in Run1 and SSP $^{232}$Th are pulled by 1.8$\sigma$ and 1.2$\sigma$, respectively, relative to their input values.
The bias of the overall efficiency obtained from the fit is $3.9\% \pm 0.6\%$ (Run0) and $-0.4\% \pm 1.4\%$ (Run1), and the bias of the signal selection is $-0.9\% \pm 1.9\%$ (Run0) and $1.2\% \pm 2.1\%$ (Run1).
The half-life of $^{136}$Xe $2\nu\beta\beta$ is measured as $2.14 \pm 0.05 \, (\text{stat.} + \text{syst.}) \times 10^{21}$ years, representing the most precise result so far, with a total uncertainty smaller than that of any previous measurement.
It is worth noting that this result is slightly smaller than our previous result from Run0 data~\cite{PandaX:2022kwg}, due to the use of the updated and slightly smaller $^{136}$Xe abundance of $8.584\% \pm 0.114\%$~\cite{PandaX:SSP}, compared to the previous value of $8.857\% \pm 0.168\%$. 
The $^{136}$Xe $2\nu\beta\beta$ spectrum is treated as the background in the search for Majoron-emitting $\beta\beta$ signals, as shown later.

In the fit of $\xi_{31}^{2\nu}$, we allow both the half-life of $^{136}$Xe $2\nu\beta\beta$ and $\xi_{31}^{2\nu}$ (i.e., the shape of $^{136}$Xe $2\nu\beta\beta$) to float freely. 
The likelihood is almost the same as Eq.~(\ref{eq:likelihood}), but with a different definition of $N_i$ in Run-$r$, which is modified as
\begin{equation}
\begin{aligned}
 N_{i} = & (1+\eta_{a}) [ (1+\eta_s) n_{s} \frac{S_{i, G0}+\xi S_{i, G2}}{1+\xi} \\
 & + \sum_{b=1}^{N_\mathrm{bkg}} (1+\eta_b) n_{b} B_{b, i}],
\end{aligned}
\end{equation}
where $S_{i, G0}$ and $S_{i, G2}$ denote the normalized energy spectrum values in the $i{\mathrm{th}}$ bin for the first and second terms of the Taylor expansion of the $2\nu\beta\beta$ decay rate, respectively.
The parameter $\xi$ is defined as $\xi = \frac{G_{2}^{2\nu}}{G_{0}^{2\nu}} \frac{\epsilon_{2}}{\epsilon_{0}} \xi_{31}^{2\nu}$, where $\epsilon_{0}$ and $\epsilon_{2}$ denote the total detection efficiencies for the theoretical spectra of the first and second terms of the Taylor expansion, respectively.
The rest of the parameters remain the same as in Eq.~(\ref{eq:exp_count}).

The backgrounds, also summarized in Table~\ref{tab:bkg_summary}, are consistent with expectations, except for the contributions from $^{\textrm{85}}$Kr in Run1 and SSP $^{232}$Th, which have pulls of 1.6$\sigma$ and 1.2$\sigma$, respectively.
The bias of the overall efficiency obtained from the fit is $3.7\% \pm 0.6\%$ (Run0) and $-0.7\% \pm 1.4\%$ (Run1), and the bias of the signal selection is $-0.9\% \pm 1.9\%$ (Run0) and $1.3\% \pm 2.1\%$ (Run1).
We obtained a best fit of $\xi_{31}^{2\nu} = 0.59^{+0.41}_{-0.38}$.

\begin{figure}[tbp]
    \includegraphics[width=1.\columnwidth]{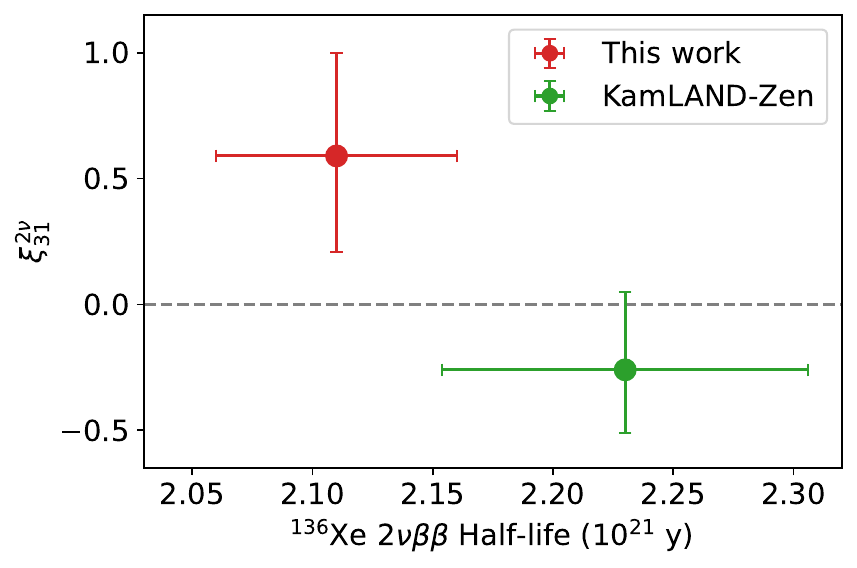}
    \caption{
 Comparison of $^{136}$Xe $2\nu\beta\beta$ half-lives and $\xi_{31}^{2\nu}$ values from the KamLAND-Zen experiment~\cite{KamLAND-Zen:kesi31} and our present analysis.
 The gray dashed line indicates the zero $\xi_{31}^{2\nu}$.
 }
    \label{fig:NME_result}
\end{figure}

Figure~\ref{fig:NME_result} compares the measured values of $\xi_{31}^{2\nu}$ and the $^{136}$Xe $2\nu\beta\beta$ half-life obtained by KamLAND-Zen~\cite{KamLAND-Zen:kesi31} and by this work. 
Although both results are consistent with $\xi_{31}^{2\nu}=0$, our measured value has the opposite sign to that reported by KamLAND-Zen ($-0.26^{+0.31}_{-0.25}$). 
Assuming fully uncorrelated uncertainties, the two measurements differ by approximately $1.7\sigma$.

Our result is consistent with quasiparticle random-phase approximation predictions~\cite{KamLAND-Zen:kesi31} at the $1\sigma$ level and remains compatible with shell-model calculations~\cite{KamLAND-Zen:kesi31} within $2\sigma$. 
However, the current precision does not allow us to exclude $\xi_{31}^{2\nu}=0$ conclusively and prevents us from discriminating between the higher-state dominance and single-state dominance hypotheses~\cite{kesi31:theory}. 
Achieving such discrimination will require improved precision in future experiments.

The inclusion of the second-order contribution is found to have a negligible impact on the extracted $^{136}$Xe $2\nu\beta\beta$ decay rate. 
The corresponding half-life is determined to be $2.11 \pm 0.05 \,(\text{stat.} + \text{syst.}) \times 10^{21}$ years, which is consistent with the result obtained from a $^{136}$Xe $2\nu\beta\beta$-only fit that neglects the second-order contribution.

\begin{figure}[tbp]
    \includegraphics[width=1.\columnwidth]{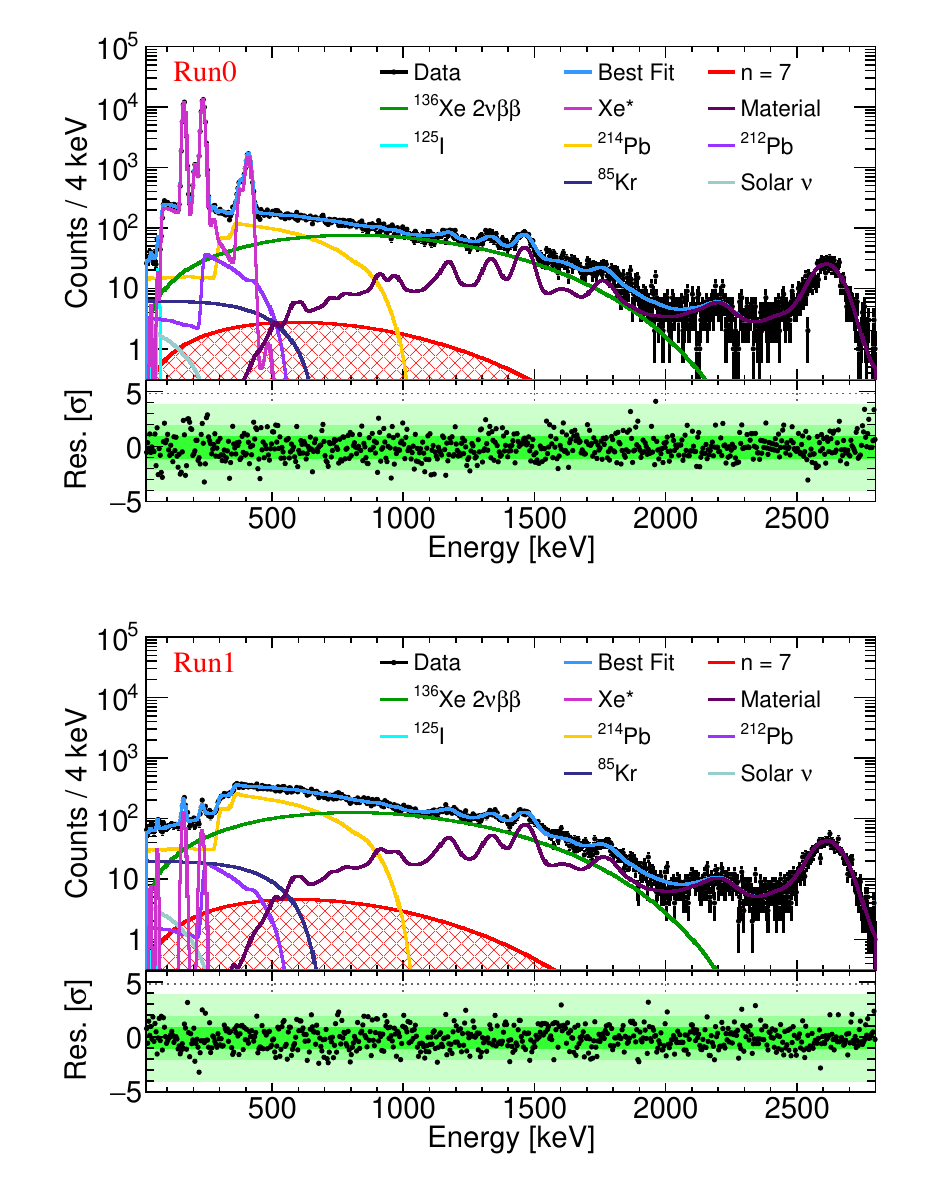}
    \caption{Spectra fits with Majoron-emitting $\beta\beta$ decay signals ($n=7$) for Run0 (top) and Run1 (bottom), respectively.
    The figures are similar to Fig.~\ref{fig:fitresult_DBD} but with the upper limit (90\% CL) on the Majoron-emitting $\beta\beta$ signal (mode $n=7$) illustrated as hatched histograms. The lower panel shows the residuals together with $\pm 1 \sigma$, $\pm 2 \sigma$, and $\pm 4 \sigma$ bands.
 }
    \label{fig:fitresult_Majoron}
\end{figure}

In the fit of Majoron-emitting $\beta\beta$, the likelihood is similar to Eq.~(\ref{eq:likelihood}), except that $^{136}$Xe $2\nu\beta\beta$ is moved from the signal term to the background term, and then, Majoron-emitting $\beta\beta$ is introduced as the signal.
As an example, the best-fit spectrum for mode $n=7$ is shown in Fig.~\ref{fig:fitresult_Majoron}. 
No statistically significant evidence for Majoron-emitting $\beta\beta$ decays is observed for any mode considered here, and the lower limits on Majoron-emitting $\beta\beta$ half-lives are derived at the $90\%$ confidence level (CL), as summarized in Table~\ref{tab:Majorons_limits}.
The extension of the lower bound of the ROI from 600~keV (as in EXO-200~\cite{EXO-200:Majoron2021}) to 20~keV considerably improves the sensitivity to mode $n=7$, where the Majoron-emitting $\beta\beta$ spectrum peaks at approximately 600~keV, leading to the most stringent constraint.
However, our sensitivity to modes $n=$ 1, 2, and 3 is limited due to our $^{136}$Xe exposure being only 0.16 times that of EXO-200~\cite{EXO-200:Majoron2021} and our higher background level.
Similarly, the backgrounds are consistent with expectations, with only the contributions from $^{\textrm{85}}$Kr in Run1 and SSP $^{232}$Th pulled by 1.7$\sigma$ and 1.2$\sigma$, respectively (Table~\ref{tab:bkg_summary}).
The bias of the overall efficiency obtained from the fit is $2.6\% \pm 1.8\%$ (Run0) and $-0.6\% \pm 1.8\%$ (Run1), and the bias of the signal selection is $-0.1\% \pm 2.1\%$ (Run0) and $0.0\% \pm 2.4\%$ (Run1).
\begin{table}[tbp]
    \caption{The 90\% CL lower limits on half-lives for different Majoron-emitting $\beta\beta$ decay models and the comparison with EXO-200 results~\cite{EXO-200:Majoron2021}}
    \label{tab:Majorons_limits}
    \centering
    \renewcommand{\arraystretch}{1.3}
    \begin{tabular}{>{\centering\arraybackslash}m{2cm}>{\centering\arraybackslash}m{3cm}>{\centering\arraybackslash}m{3cm}}
    \toprule
    \multicolumn{1}{c}{Decay mode} & \multicolumn{1}{c}{This work (yr)} & \multicolumn{1}{c}{EXO-200 (2021) (yr)} \\
    \midrule
    $n$ = 1 & $\textgreater  2.4 \times 10^{23}$ & $\textgreater 4.3 \times 10^{24}$ \\
    $n$ = 2 & $\textgreater 1.2 \times 10^{23}$ & $\textgreater 1.5 \times 10^{24}$ \\
    $n$ = 3 & $\textgreater 5.7 \times 10^{22}$ & $\textgreater 6.3 \times 10^{23}$ \\
    $n$ = 7 & $\textgreater 7.0 \times 10^{22}$ & $\textgreater 5.1 \times 10^{22}$ \\
     \bottomrule
    \end{tabular}
\end{table}

In summary, we perform a measurement of $\xi_{31}^{2\nu}$ in $2\nu\beta\beta$ and a search for Majoron-emitting $\beta\beta$ for $^{136}$Xe with a total $^{136}$Xe exposure of $39.1 \pm 0.7$~kg$\cdot$yr using PandaX-4T data. 
The most precise measurement of the $^{136}$Xe $2\nu\beta\beta$ half-life to date is achieved thanks to the nearly complete spectrum used in the fit.
Our measurement of $\xi_{31}^{2\nu}$ is consistent with theoretical predictions. 
The $\xi_{31}^{2\nu}$ values from KamLAND-Zen and this work are broadly consistent, with a modest ($\sim$1.7$\sigma$) difference.
The most competitive results for the Majoron-emitting $\beta\beta$ mode with $n=7$ are obtained.
Following detector upgrades in 2023, PandaX-4T has been taking data in the second science run (Run2).
The capability to reconstruct the nearly complete $^{136}$Xe $\beta\beta$ spectrum will allow us to fully exploit the PandaX-4T dataset, significantly enhancing its physics potential.

\nolinenumbers

% !TEX root = ../main.

%\section{Acknowledgement}

\textit{Acknowledgements}\textemdash This project is supported in part by grants from the National Key R\&D Program of China (No. 2023YFA1606200 and No. 2023YFA1606202), National Science Foundation of China (No. 12090060, No. 12090062, No. 12305121, No. U23B2070), and by Office of Science and Technology, Shanghai Municipal Government (Grants No. 21TQ1400218, No. 22JC1410100, No. 23JC1410200, No. ZJ2023-ZD-003). We are grateful for the support of the Fundamental Research Funds for the Central Universities. We also thank for their sponsorship: the Chinese Academy of Sciences Center for Excellence in Particle Physics (CCEPP), Thomas and Linda Lau Family Foundation, New Cornerstone Science Foundation, Tencent Foundation in China, and Yangyang Development Fund. Finally, we thank the CJPL administration and the Yalong River Hydropower Development Company Ltd. for indispensable logistical support and other help.

%\section{Data availability}

\textit{Data availability}\textemdash The data that support the findings of this article are not publicly available upon publication because it is not technically feasible and/or the cost of preparing, depositing, and hosting the data would be prohibitive within the terms of this research project. The data are available from the authors upon reasonable request.
\bibliographystyle{apsrev4-2}
\bibliography{P4NMEMajorons}
\end{document}